\documentstyle[12pt]{article}
\begin{document}
\begin{center}
{\Large \bf Induced Chern-Simons term in lattice QCD at finite temperature}\\
\vspace{1cm}
{\large O.A.~Borisenko, V.K.~Petrov,
G.M.~Zinovjev\footnote{email: gezin@gluk.apc.org}}\\[.5cm]
{\large \it
N.N.Bogolyubov Institute for Theoretical Physics, \\ 252143 Kiev,
Ukraine}
\end{center}
\vspace{.5cm}
\begin{abstract}
The general conditions for the Chern-Simons action to be
induced as a nonuniversal contribution of fermionic determinant are formulated
in the finite temperature lattice QCD. The dependence of
the corresponding action coefficient on nonuniversal parameters (chemical
potentials, vacuum features, etc. ) is explored. Special attention
is paid to the role of $A_0$-condensate if it is available in this
theory.
\end{abstract}

\newpage

\section{Introduction and Motivation}

The principle that gauge invariance demands massless gauge fields
has produced far reaching consequences in so many branches of
physics that possible peaceful coexistence of gauge invariance and
massive gauge fields has been just recently established \cite{1}. There
exists a mass term for gauge field Lagrangian in $(2+1)$-dimensions
that keeps the gauge invariance and for nonabelian theories exhibits
profound topological meaning, so-called the Chern-Simons (CS) term \cite{2}.
Thereby the initial gauge field Lagrangian should be
\begin{equation}
L = L_{G} + L_{CS}
\label{ltot}
\end{equation}
\noindent
where  $L_G$ is the usual Maxwell or Yang-Mills term and  $L_{CS}$ is the
term providing a mass for the gauge field. For nonabelian  gauge
group
\begin{equation}
L_{CS} = -k\epsilon^{lmn}Tr[\partial_l A_m A_n + \frac{2}{3}A_l A_m A_n]
\label{lcs}
\end{equation}
\noindent
with  $\epsilon^{lmn}$ as the totally antisymmetric tensor and the
constant $k$ must have a dimensionality of mass. (The Latin indices are used
for 3-dimensional theory).
Although the equations of motion are gauge invariant, $L_{CS}$ is
not and transforms under large gauge transformations  $g$ with
\begin{equation}
A_m \rightarrow g^{-1}A_m g + g^{-1} \partial_m g
\label{gtr}
\end{equation}
\noindent
leading to
\begin{eqnarray}
L_{CS} \rightarrow L_{CS} + k \epsilon^{lmn} \partial_l (Tr
[\partial_m g g^{-1} A_n]) + 8\pi^{2}kw(g)  \nonumber  \\
w(g) = (24\pi^{2})^{-1}\epsilon^{lmn}Tr
[g^{-1}\partial_l g g^{-1}\partial_m g g^{-1}\partial_n g]
\label{lcstr}
\end{eqnarray}
\noindent
Here all change has to be a total derivative, which is valid only
locally in group space for $w(g)$. Then the relevant action $S_{CS}=
\int d^{3}x L_{CS}$ includes the part proportional to the winding number
\begin{equation}
W(g) = \int d^{3}x w(g)
\label{wn}
\end{equation}
which is an integer for compact nonabelian gauge group. This is
the origin of $k$ quantization. The complete analysis (it can be
fulfilled properly for the abelian theory) establishes that $k$ gives
a mass for the excitations and we have massive gauge field
which nevertheless respects gauge invariance. Moreover, unlike the
Higgs mechanism, the presence of the CS mass term does not change
the number of physical degrees of freedom for gauge field and the
CS term does not get the quantum corrections for topological
reason in non-abelian theory \cite{3}. It has been also shown \cite{4} that
in the theory of gauge fields interacting with fermions the
topological mass term can be induced by such an interaction and
long-distance effective gauge theory includes the CS term.

   Initially this construction has arisen to get insight into
quantum field theory. However, when the quantum field theory at
finite temperature has been recognized to become effectively
three-dimensional \cite{5}, \cite{6} an idea of possible physical applications
of the CS theory has appeared. In particular, it has been conjectured
that the CS term could play an important role in regulating the infrared
behaviour of QCD high temperature phase \cite{7}. As is known the fermion
and nonstatic boson modes have been seen in perturbation theory to
acquire a mass proportional to the temperature and one
has believed that in the effective three-dimensional thermal
theory the CS term should produce magnetic mass to screen  static
gauge (magnetic) field. But this highly  desirable conclusion  has
not been held after gazing at topologically massive chromodynamics
at finite temperature \cite{8}, \cite{9}.  Nevertheless the interest in the
CS theory has been warmed up by recent attempts to explain
the quantum Hall effect \cite{10} and high-$T_{c}$ superconductivity
\cite{11} drawing planar gauge theory dynamics.

Very important observation in this context is that the
Chern-Simons term violates $P$ and $T$ symmetry.
When the CS term is induced by fermion radiative
corrections \cite{4} the Fermi fields in three-dimensional space-time  are
described by two-component spinors and the fermion mass term
violates $P$ and $T$. It is also justified for massless
fermions due to the mass term emerging in Pauli-Villars
regularization which is necessary to keep gauge invariance against
small gauge transformations with zero winding number. A
comprehensive analysis of the origin of the induced Chern-Simons term
makes one to recall that the complicated vacuum structure ($\theta$-vacuum)
of QCD results in the prediction of $P$ and $T$-violation
in strong interactions unless $\theta = 0$. As is well-known, the term
\begin{equation}
L_{i\theta} = -\frac{\theta g^{2}}{32\pi^{2}} \tilde{G}G
\label{ldual}
\end{equation}
\noindent
where $\tilde{G}$ is the dual tensor, being added to the complete QCD action
\begin{equation}
S_{QCD} = \int_{\Omega}[dx] (L_G + L_F)
\label{stot}
\end{equation}
\noindent
with a fermion Lagrangian $L_F$, is the only one  preserving
the QCD Lagrangian as gauge invariant and  renormalizable.
Being  the 4-divergency this term (\ref{ldual}) does not lead to any  change
of the equations of motion. In Eq.(\ref{stot}) $\Omega$ means various
possibilities, in particular,
\begin{equation}
\Omega = (R^{4};\ R^{3}\otimes S^{1};\ S^{1}\otimes S^{1} \otimes S^{1}
\otimes S^{1} = T^{4}).
\label{space}
\end{equation}
\noindent
Integration over $R^{4}$ denotes the usual space-time consideration,
the finite
temperature regime leads to the $\Omega = R^{3} \otimes S^{1}$
integration when the periodic or other boundary conditions are settled in
one direction and the space $\Omega = T^{4}$  appears, in fact, at   Monte
Carlo lattice calculations. The present experimental estimate of
$\theta$ ranging up to $\theta \leq 10^{-9}$ \cite{12} allows us formally to
put
$\theta = 0$. As a spontaneous breaking of $P$ symmetry is banned by the
Vafa-Witten theorem \cite{13}, $P$ and $T$-symmetry violations
could  be  caused for $\Omega = R^{4}$-space
by certain  exotic mechanism.

The  situation  is  more  sophisticated  and  promising  for
$\Omega = {R^{3}\otimes S^{1};T^{4}}$ in Eq.(\ref{stot}) with
nonzero baryonic chemical potential $\mu_{B} \neq 0$.
There are enough reasons to believe the Vafa-Witten theorem is
inapplicable for this approach. The
proof of the theorem  is based upon the fact that any $P$-
nonsymmetric operator $Q$ (if it is vector or tensor) implanted into
a functional integral with a corresponding source in Euclidean
space (just after the Wick rotation) becomes pure imaginary.
Then, the ground state will always be upper than the state where
$<Q> \neq 0$. However, the presence of baryonic chemical potential
makes the action, from the beginning, a complex quantity,
thereby breaking the basic conditions of the theorem. In particular,
it has been shown for scrutinized case that for $SU(2)$-theory
in the strong coupling regime a baryonic condensate falls
down leading to spontaneous breaking of  $U(1)$ vectorial symmetry
\cite{14}.The principal conditions of the Vafa-Witten theorem are not
fulfilled for gauge theories on torus. When the gauge
field winds a torus it changes up to gauge transformation
\cite{15}. The corresponding matrices should obey special requirement
of selfconsistency. For example, the transformations of Cartan
subgroup are admissible. At such constraints the gauge field
matrices should gain complex factors and belong to various
topological classes to be finally summed up over \cite{16}.

Ultimately, the status of that theorem is still unclear for the
theory with singular potentials \cite{17} as well as for the theory in
a dielectric vacuum which is generally described for $SU(3)$-theory
by the complex matrices \cite{18}. For both cases the fermionic
determinant is not a positively defined quantity and has to
contain imaginary additions being  at variance with the
conditions of the Vafa-Witten theorem \cite{13}. Inapplicability of
the latter in any case does not, certainly, mean that $P$ or
$CP$-symmetries should be spontaneously broken.

If the $\theta$-term is out of in Eq.(\ref{stot}) the
only way to extract the CS term is to  explore an
effective action generated by fermions coupled to a nonabelian
gauge field in three-dimensional Euclidean space \cite{4}. This action
results after the integration over Fermi-fields giving for
the partition function
\begin{equation}
Z = \int [dA_{\mu}] \exp [-\int_{\Omega} [dx] L_G - \Gamma_{eff}(A)]
\label{ztot}
\end{equation}
\noindent
with $\Gamma_{eff}(A)$ being the logarithm of the Dirac operator determinant
\begin{equation}
\Gamma_{eff}(A) = - \ln detM(A)
\label{geff}
\end{equation}
\noindent
where $M(A) = D + m$  with covariant derivative $D$ and fermion mass $m$ of
one flavour for simplicity. The conventional method of studying $CP$-
or $P$-odd effects is to consider the expansion of the effective action
in powers of the gauge field.  One  normally  encounters  the
same conclusion as to $P$- or $CP$-odd effects  independently  of the
method of determinant analysis. They result from the Chern-Simons
action \cite{22}, \cite{23} though the coefficients of the Chern-Simons term
turn out to be different for various regularization schemes \cite{24}.

In what follows we explore the possible situation when the Vafa-Witten
theorem is plausibly inapplicable.  First of all,
analyzing a possibility of $C$ or $CP$-symmetry breaking caused by
the generation of $A_0$-condensate \cite{19} and/or the baryonic
chemical potential $\mu_{B}$ we clarify
any relation of this phenomenon in finite temperature QCD with the
Chern-Simons action. Moreover, $A_0$-condensate should be taken as
playing a role of imaginary chemical potential of colour quark charges.
It has been argued in \cite{20} that the global
gauge symmetry is also lost at the deconfining phase transition
and it has been shown by means of Monte Carlo simulations for lattice
gluodynamics in unitary gauge that in the high temperature phase
a nonzero expectation value of gauge field component in compactified
direction $<A_0>$ is developed. Detailed studying of the effective potential
generated by quantum fluctuations has displayed the  breakdown  of
initial SU(3)-symmetry up to its Abelian subgroup  $U(1) \otimes U(1)$ at
the deconfinement temperature. Since finite temperature
consideration leads to  compactification in the imaginary  time
direction an effective theory invariant with respect to the Cartan
subgroup should be three-dimensional \cite{6}, \cite{21} and could include  the
Chern-Simons term \cite{4} justifying the way of inducing this term.

Secondly, the discrete symmetry breaking manifests itself within certain
regularization
schemes. In particular, $CP$ or $P$-symmetry breaking survive even after
removing the regularization. Then the fermionic determinant can generate
either $\theta$-term or CS-term in the corresponding limit (see, for instance,
\cite{24}). We go through this item combining various chemical potentials
($\mu_{B}$, $<A_{0}>$) with some vacuum features (for example, non-zero
dielectric constant).

In order to keep a nonperturbative information as far as
possible  we are dealing with the lattice regularization of
gauge theories in what follows. We are not solving the problem of
introducing a lattice  analogy of the Chern-Simons interaction
directly. It is still the pending question because of large ambiguity
in the definition of the wedge product \cite{25}. The final goal of our
exploration is to formulate the most general features of $\Gamma_{eff}(A)$
when the coefficient of the Chern-Simons action $S_{CS}$ comes about to
be nonzero. There exist many ways to put down the fermions
on a lattice. We discuss here only two of those making sure
that the generation of the Chern-Simons term is strongly dependent on
a concrete form of lattice fermion action.

The  organization of our paper is as follows. First, in Sec.2, we
are constructing the perturbative expansion of $\Gamma_{eff}(A)$  employing
the Wilson fermions with no fermion doubling. Executing the reduction
to the three dimensional formulation we find out the conditions
when the coefficient of $S_{CS}$  is non-zero in this approach. Sec.3 is
devoted to studying the gauge theory on a torus with
the Kogut-Susskind fermions. Analyzing the four-dimensional
formulation we deduce the properties of $\Gamma_{eff}(A)$  when  it includes
the $\theta$-term. Then  it is shown how the latter is reduced to  $S_{CS}$
when  $<A_0 > \neq 0$  is available. Finally, the conclusions are drawn in
Sec.4 to summarize the necessary and sufficient conditions to include
$S_{CS}$.

\section{Generalized Wilson fermions}

Since a naive rewriting of the lattice fermion
action generates  the problem of fermion doubling \cite{26} there exist two
widely used forms of the lattice fermion action: the Wilson action
$S^{F}_W$ and the Kogut-Susskind action $S^{F}_{KS}$ where this
problem  is solved completely (for $S^{F}_W$ ) or partially (for $S^{F}_{KS}$).
However, it leads to the breakdown of some symmetries of the Lagrangian
at the classical level (it disappears in the continuum limit).
We  start from the  Wilson action
\begin {eqnarray}
S_W &=& \frac{1}{2}a^{3}\sum_{x,\mu}[\overline{\Psi}(x)U_{\mu}(x)
(r-\gamma_{\mu})
\Psi(x+a_{\mu}) + \overline{\Psi}(x)U_{\mu}^{+}(x-a_{\mu})
(r+\gamma_{\mu})\Psi(x-a_{\mu})] \nonumber \\
 &+& a^{4}m\sum_{x}\overline{\Psi}(x)\Psi(x) - da^{3}\sum_{x}
\overline{\Psi}(x)R\Psi(x).
\label{Wact}
\end {eqnarray}
\noindent
It was originally taken with $r=1$ \cite{27}.  Then it has soon become clear
\cite{28} that the more general choice $r=s\exp (i\theta \gamma_5)$
with $0 < s \leq 1$  is admissible. In fact, the naive continuum limit
takes place at any set of parameters and leads to the theory of Dirac
massive fermions coupled to a smooth gauge field, but the chiral
symmetry is explicitly broken at any $\theta$ and $s$, and  for
$\theta \neq 0;\pi$ the CP-symmetry is broken as well.
The Green function takes the form  (at  zero temperature)
\begin{equation}
G^{-1}_{\theta ,s}(p)=\frac{s}{a} \exp (i\theta \gamma_5)
\sum_{\mu}(\cos ap_{\mu}-1) + \frac{\gamma^{\mu}}{a}\sin ap_{\mu} + m
\label{grf1}
\end{equation}
\noindent
where it is quite transparent that the elimination of unnecessary
degrees of freedom happens as well as for $r=1$. However, $\theta$
dependence persists in the quantum theory even after taking the
continuum limit. In Ref.\cite{29} the result was obtained
\begin{equation}
\lim_{a \rightarrow 0} \frac{Z_{\theta}(A)}{Z_{\theta}(0)} =
e^{i \theta Q} \lim_{a \rightarrow 0} \frac{Z_0(A)}{Z_0(0)}
\label{qlim}
\end{equation}
where
$$
Z_{\theta}(A)= \int D \bar{\Psi}D\Psi \exp [-S_W(\theta)]
$$
and $Q$ is defined as
$$
Q= \frac{1}{32 \pi^{2}} \epsilon^{\mu \nu \alpha \beta}
Sp (F_{\mu \nu} F_{\alpha \beta})
$$
Thus, despite the classical limit is $\theta$-independent and $CP$-violation
is absent, it does not take place in the quantum theory. Important
result for our motivation comes also from Ref.\cite{24} where
three-dimensional Abelian theory was considered for  $r=-1$ with
the action (\ref{Wact}) and the following Green function
\begin{equation}
G^{-1}(p) = m - \frac{i}{a}\sigma_n \sin ap_n +
\frac{2r}{a} \sin^{2}\frac{ap_n}{2}
\label{grf2}
\end{equation}
\noindent
where $\sigma_n$ are the Pauli matrices. It follows from  Eq.(\ref{grf2})
that there is no species doubling, the propagator is regular at all
$p \neq 0$  and is reduced in the continuum limit to the standard Feynman
form. Moreover, a positive transfer matrix acting  in  the  proper
Hilbert space does exist. One could conclude that the lattice theories
with $r = \pm 1$  are in the same universality  class. However,
it is not true even in perturbation theory. The  results  of
Ref.\cite{24} show that the fermionic determinant at $r=-1$ generates the
Chern-Simons term  $S_{CS}$ and in the continuum  limit   we  have for
effective action
\begin{eqnarray}
\lim_{m \rightarrow 0}Im \Gamma (A) &=& c_0 g^{2}S_{CS} + \pi h(A),
\nonumber   \\
\lim_{m \rightarrow \infty} \Gamma (A) &=& ic_{\infty}g^{2}S_{CS},
\label{efactl}
\end{eqnarray}
here $h(A)$ is an integer and $c_0 = c_{\infty}+\pi$,$c_{\infty}=2\pi n$.
We could,
thus, make a conclusion that the Wilson parameter $r$  plays an
important role of the quantity dividing the initial action into
different  universality classes that are characterized by the
parameters of the nonuniversal terms of an effective action ($\theta$-term or
Chern-Simons term) and are defined by the parameter $r$.  The
calculation  we provide here confirms this assumption.

{}From now on we use for the parameter $r$ the most general expression of
Ref.\cite{30}
\begin {equation}
r = s\exp(\imath \theta \gamma_{5})T,
\ 0 < s \leq 1, \ 0\leq \theta \leq \pi,
\label{Wpar}
\end {equation}
\noindent
where $T$ is an arbitrary hermitian matrix acting in fermion space and
$T^{2}=1$ should be valid. If  $T=\gamma_{\mu}$ the action at $m=0$ is
chiral symmetric and at arbitrary $T$  and $m=0$ it is invariant under
transformations \cite{30}
\begin{equation}
\Psi \rightarrow \exp [i \alpha \gamma_5 \tilde{T}] \Psi , \
\bar{\Psi} \rightarrow \bar{\Psi} \exp [i \alpha \gamma_5 \tilde{T}]
\label{cgsym}
\end{equation}
\noindent
where $\tilde{T}$ anticommutes with $T$. We  are  interested  here  in
$T=\gamma_0 T_l$ ( $T_l$ is acting, for example, in flavour space) as it
leads to the Chern-Simons term  ( of course, the other possibilities
are not excluded). Denoting lattice spacing in spatial (temperature)
direction as $a_{\sigma}$ ($a_{\beta}$) and space dimension as $d$, we
have for the action
\begin {eqnarray}
S_W &=& \frac{1}{2}a_{\sigma}^{2}a_{\beta}\sum_{x} \sum_{\sigma=1}^{d}
[\bar{\Psi}(x)U_{\sigma}(x) \eta^{-}_{\sigma} \Psi(x+a_{\sigma})+
\bar{\Psi}(x)U_{\sigma}^{+}(x-a_{\sigma}) \eta^{+}_{\sigma}\Psi(x-a_{\sigma}]
\nonumber \\
 &+& \frac{1}{2}a_{\sigma}^{3}\sum_{x}
[\bar{\Psi}(x)U_0(x) \eta^{-}_0 \Psi(x+a_{\beta}) +
\bar{\Psi}(x)U_0^{+}(x-a_{\beta}) \eta^{+}_0 \Psi(x-a_{\beta})] \nonumber  \\
 &+& m a_{\sigma}^{3}a_{\beta} \sum_{x} \bar{\Psi}(x) \Psi(x) -
d a_{\sigma}^{3}a_{\beta} s \sum_{x} \bar{\Psi}(x)
\exp(i \theta \gamma_{5})T \Psi(x)  \nonumber  \\
 &-& a_{\sigma}^{3} s \sum_{x} \bar{\Psi}(x)
\exp(i \theta \gamma_{5})T \Psi(x).
\label{Wact1}
\end {eqnarray}
\noindent
We introduced here
\begin{eqnarray}
\eta^{\pm}_{\sigma} &=& \exp [\pm a_{\sigma} \alpha]
(\exp(i \theta \gamma_{5})T \pm \gamma_{\sigma})  \nonumber   \\
\eta^{\pm}_0 &=& \exp [\pm a_{\beta} \mu k]
(\exp(i \theta \gamma_{5})T \pm \gamma_0) \exp (\mp i \Lambda)
\label{eta}
\end{eqnarray}
\noindent
where $\gamma_{\mu}$ represents Euclidean version of the Dirac
$\gamma$-matrices,
$\alpha$ can be interpreted here as a chemical potential if it is real or
as an external constant Abelian field if it is imaginary and $\mu$ is the
baryonic chemical potential. Then $k=1$ if this chemical potential is
conventionally introduced on the lattice \cite{31} or $k=\gamma_5$ if
we introduce the chemical potential similarly to \cite{32} having different
chemical potentials for the left- and right-hand fermions, i.e. $\mu k$ is
the matrix
\begin{equation}
\left( \begin{array}{lllll}
 \mu_1	& 0 &     \\
 0	& \mu_2  &     \\
\end{array} \right)
\label{chempot}
\end{equation}
\noindent
with $\mu$ being 2x2 matrices, and finally $\Lambda=<a_{\beta}gA_0>$.
We do not fix the gauge here assuming periodic boundary
conditions for the gauge field and antiperiodic ones for the
fermion field. Constructing a perturbative expansion of the fermionic
determinant, we keep in mind only definitions (\ref{ltot}) and (\ref{lcs}) as
any such expansion for the smooth $A_{\mu}(x)$ potentials brings us
directly to the continuum. This remark is rather important as lattice
non-abelian formulation of $S_{CS}$ is not yet well developed. A
finite-difference form for that is not gauge-invariant and using
the compact lattice fields we come to the same problems as for the $\theta$-
term \cite{33} (its standard definition on a lattice is correct only in
the week coupling region). For Abelian lattice theory this problem
is hopefully solved and $S_{CS}$ can be given in the gauge invariant
form \cite{33}.

In order to construct the necessary expansion for $g \approx 0$ we
introduce for the gauge field matrices
\begin{equation}
U_{\mu}(x) = \sum_{k=0}\frac{(i a_{\mu}g)^{k}}{k!} [A_{\mu}(x)]^{k}
\label{uexp}
\end{equation}
\noindent
($A_{\mu}(x)=A_{\mu}^{c}(x) t^{c}$, where $t^{c}$ are the generators of
$SU(N)$ in the fundamental representation).
Then the effective action in (\ref{geff}) can be easily given as
\begin{equation}
\Gamma_{eff}(A) = -Sp \ln [G+\sum_{k=1}g^{k}D_k]
\label{geff1}
\end{equation}
\noindent
and
\begin{eqnarray}
G = \frac{\xi}{2}\sum_{\sigma=1}^{d}(\eta^{-}_{\sigma}\delta^{x+a_{\sigma}}_y
+ \eta^{+}_{\sigma}\delta^{x-a_{\sigma}}_y) +\frac{1}{2}
(\eta^{-}_0 \delta^{x+a_{\beta}}_y + \eta^{+}_0 \delta^{x-a_{\beta}}_y) +
\delta_y^{x} (ma_{\beta} - d\xi r -r) \ , \nonumber   \\
D_k = \frac{\xi}{2} \sum_{\sigma=1}^{d}[(ia_{\sigma}A_{\sigma}(x))^{k}
\frac{1}{k!}\eta^{-}_{\sigma}\delta^{x+a_{\sigma}}_y  +
(-ia_{\sigma}A_{\sigma}(x))^{k}
\frac{1}{k!}\eta^{+}_{\sigma}\delta^{x-a_{\sigma}}_y ] +
\label{grf3}
\end{eqnarray}
\noindent
$$
\frac{1}{2}(ia_{\beta}A_0(x))^{k}
\frac{1}{k!}\eta^{-}_0 \delta^{x+a_{\beta}}_y  +
\frac{1}{2}(-ia_{\beta}A_0(x))^{k}
\frac{1}{k!}\eta^{+}_0 \delta^{x-a_{\beta}}_y
$$
with $\xi = \frac{a_{\beta}}{a_{\sigma}}$ or, rewriting, we obtain
\begin{equation}
\Gamma_{eff}(A) = -\sum_{l=1}^{\infty}\frac{(-1)^{l}}{l}
Sp[(\sum_{k=1}g^{k}D_k)G^{-1}]^{l} -Sp \ln G
\label{geff2}
\end{equation}
\noindent
It is clear now that only $D_1$ contributes to $S_{CS}$ as $I(A)$ is a
third power polynomial in $A_{\mu}(x)$ and it includes various components
of the potential $A_{\mu}(x)$. It brings us to
\begin{equation}
\Gamma_{eff}(A) = \frac{g^{2}}{2}[\Gamma_2(A) -\frac{2}{3}g\Gamma_3(A)] -
\tilde{\Gamma}_{eff}(A)
\label{geff3}
\end{equation}
\noindent
\begin{equation}
\Gamma_{k}(A) = Sp [(D_1G^{-1})^{k}]
\label{gk}
\end{equation}
\noindent
In Eq.(\ref{geff3}) $\tilde{\Gamma}_{eff}(A)$ includes the higher
contributions of $D_1$  together with contributions  of $D_k(k>2)$,
exhibiting probably universal behaviour as will be  discussed below.
Now, Fourier-transforming the gauge potentials
\begin{equation}
A^{a}_{\mu}(\vec{x}, \tau) = \int_{-\pi /a_{\sigma}}^{\pi /a_{\sigma}}
\frac{d^{3}q}{(2\pi)^{3}}\sum_{n=\frac{N_{\beta}}{2}-1}^{\frac{N_{\beta}}{2}}
\exp [iq_{\sigma}(x_{\sigma}+\frac{1}{2}a_{\sigma}) + \frac{2\pi i}{N_{\beta}}
n\tau] \tilde{A}^{a}_{\mu}(q_{\sigma},n)
\label{fuortr}
\end{equation}
\noindent
(time periodicity is taken into account here) we have for the Green
function (making allowance for fermion antiperiodicity)
\begin{eqnarray}
\tilde{G}(p,k) = S^{-1} = \frac{1}{2a_{\sigma}}\sum_{\sigma=1}^{d}
(\eta^{-}_{\sigma}\exp(-ip_{\sigma}a_{\sigma}) \ +
\eta^{+}_{\sigma}\exp(ip_{\sigma}a_{\sigma}))  \ +   \nonumber   \\
\frac{1}{2a_{\beta}}(\eta^{-}_0\exp[-\frac{2\pi i}{N_{\beta}}(k+\frac{1}{2})]
+ \eta^{+}_0\exp[\frac{2\pi i}{N_{\beta}}(k+\frac{1}{2})]) + m -
\frac{dr}{a_{\sigma}} - \frac{r}{a_{\beta}}
\label{grf4}
\end{eqnarray}
\noindent
Summing up properly in Eq.(\ref{gk}) we find
\begin{equation}
\Gamma_2(A)= \int_{-\pi /a_{\sigma}}^{\pi /a_{\sigma}}
\frac{d^{3}q}{(2\pi)^{3}}\sum_{n=\frac{N_{\beta}}{2}-1}^{\frac{N_{\beta}}{2}}
\tilde{A}_{\mu}(q,n) \Pi_{\mu \nu} \tilde{A}_{\nu}(-q,-n).
\label{g2}
\end{equation}
\noindent
The exact expression for $\Pi_{\mu \nu}(q, n)$ is very unwieldy and we drop
it here as in what follows we shall need only its reduced
form. The reduction is quite standard \cite{21}, \cite{6}
and means that in all
summations over discrete frequences we keep zero modes only as
contributing dominantly at high temperatures. Moreover, we neglect the
dynamical part of $A_0$, i.e. fluctuations around $<A_0>$, which become
massive \cite{34}. This  approximation does not conflict with periodic
boundary  conditions unlike in \cite{32}  where gauge fixing
$A_0=0$ is incompatible with boundary conditions. The  dominant
contribution comes from $<A_0>$-condensate changing effectively
the vacuum which is reflected in the structure of Green functions.
Afterwards the effective action becomes
\begin{equation}
\Gamma_2(A)= \int_{-\pi /a_{\sigma}}^{\pi /a_{\sigma}}
\frac{d^{3}q}{(2\pi)^{3}}
\bar{A}^{a}_n(q) \bar{\Pi}_{n,m}^{ab}(q) \bar{A}_m^{b}(-q).
\label{g2eff}
\end{equation}
\noindent
where all bars mean three-dimensional quantities and  the  reduced
tensor $\bar{\Pi}^{ab}_{nm}$ takes the form
$$
\bar{\Pi}^{ab}_{nm} = Sp_c Sp_s \int_{-\pi /a_{\sigma}}^{\pi /a_{\sigma}}
\frac{d^{3}q}{(2\pi)^{3}}[\eta^{-}_nt^{a}\bar{G}^{-1}(p-\frac{q}{2})
\eta^{-}_mt^{b}\bar{G}^{-1}(p+\frac{q}{2})e^{-i(p_n+p_m)} +
$$
$$
\eta^{+}_nt^{a}\bar{G}^{-1}(p-\frac{q}{2})
\eta^{+}_mt^{b}\bar{G}^{-1}(p+\frac{q}{2})e^{i(p_n+p_m)} -
\eta^{-}_nt^{a}\bar{G}^{-1}(p-\frac{q}{2})
\eta^{+}_mt^{b}\bar{G}^{-1}(p+\frac{q}{2})e^{i(p_m-p_n)} -
$$
\begin{equation}
\eta^{+}_nt^{a}\bar{G}^{-1}(p-\frac{q}{2})
\eta^{-}_mt^{b}\bar{G}^{-1}(p+\frac{q}{2})e^{-i(p_m-p_n)}]
\label{ptens}
\end{equation}
\noindent
$a$, $b$ are colour indices; $n,m = 1,2,3$, and trace is taken over the
colour (spinor) indices. In the  $x$-space this approximation implies
that we use the static configurations averaged over all time interval
\begin{equation}
\bar{A}_{\mu}(x) = \frac{1}{\beta} \int_{0}^{\beta} d \tau
A_{\mu}(x, \tau).
\label{aaver}
\end{equation}
\noindent
The reduced Green function can be utilized in the following form
\begin{eqnarray}
\bar{G}(p) = \bar{S}^{-1} =m + i\frac{\gamma_{\sigma}}{a_{\sigma}}
\sin a_{\sigma}(p_{\sigma}-i\alpha_{\sigma}) + \frac{r}{a_{\sigma}}
\sum_{\sigma=1}^{d}[\cos a_{\sigma}(p_{\sigma}-i\alpha_{\sigma}) -1] \ +
	  \nonumber   \\
\frac{1}{a_{\beta}}(\eta^{-}_0 \exp (-\frac{\pi i}{N_{\beta}})
+ \eta^{+}_0 \exp (\frac{\pi i}{N_{\beta}}) - r)
\label{grf5}
\end{eqnarray}
\noindent
($\tilde{G}(p)$ are still diagonal matrices in colour space due to the
presence of $\Lambda$). Similar but much more tiresome calculations
can be made for $\Gamma_3(A)$. As a result we have
\begin{equation}
\Gamma_3(A) = \int_{-\pi /a}^{\pi /a}\frac{d^{3}q_1d^{3}q_2}{(2\pi)^{6}}
\bar{A}_n^{a}(q_1) \bar{A}_m^{b}(q_2) \bar{A}_k^{c}(-q_1-q_2)
\bar{B}_{nmk}^{abc}(q_1,q_2).
\label{g3eff}
\end{equation}
\noindent
Now we are ready to calculate the coefficient in front of the Chern-Simons
term. Bringing the action $S_{CS}$ into the momentum representation we
have from Eq.(\ref{ltot}) and Eq.(\ref{lcs})
\begin{equation}
S_{CS} = -k\epsilon^{nmk}Sp \int \frac{d^{3}q_1}{(2\pi)^{3}}
[A_nq_m A_k + \frac{2}{3} \int \frac{d^{3}q_2}{(2\pi)^{3}} A_n(q_1)A_m(q_2)
A_k(-q_1-q_2)]
\label{csmsp}
\end{equation}
\noindent
Introducing the definition $g\overline{A} \rightarrow  \overline{A}$
into effective action and comparing $\Gamma_2(A)$ and $\Gamma_3(A)$
with (\ref{csmsp}) we  recognize that small momentum behaviour
of $\bar{\Pi}_{nm}$ and $\bar{B}_{nmk}$  is very important.
Since at  $m > 0$ the Green functions (\ref{grf5})
are regular for any $p$ we can construct their power
expansion in $p$. Then, as $\bar{\Pi}_{nm}(0) = 0$ we have
\begin{equation}
\bar{\Pi}_{nm}(q) \approx \sum_{k}\frac{\partial \bar{\Pi}_{nm}}
{\partial q_k} \mid_{q=0} \ q_k \ +O(q^{2})
\label{piexp}
\end{equation}
\noindent
and calculating
\begin{equation}
\partial_n \bar{G}(p) = i \gamma_n \cos a(p_n-i\alpha_n) -
r \sin a(p_n-i\alpha_n)
\label{grder}
\end{equation}
\noindent
(no summation over $n$) we can write
\begin{equation}
\bar{\Pi}_{nm}^{ab}(q) = -Sp_cSp_s
\int_{-\pi /a}^{\pi /a} \frac{d^{3}p}{(2\pi)^{3}}
\label{pired}
\end{equation}
\noindent
$$
[ ( \partial_n \bar{G}(p) ) t^{a} \bar{G}^{-1} (p-\frac{q}{2})
( \partial_m \bar{G}(p) ) t^{b} \bar{G}^{-1} (p+\frac{q}{2}) ]
$$
{}From (\ref{csmsp}) and (\ref{pired}) it is easy to conclude that
\begin{equation}
[\partial_k \bar{\Pi}_{nm}(q)] \mid_{q=0} = a_0 \epsilon_{nmk}
\label{pider}
\end{equation}
\noindent
where
\begin{equation}
a_0 = \frac{1}{6} \int_{-\pi /a}^{\pi /a}\frac{d^{3}p}{(2\pi)^{3}}
\epsilon_{nmk}Sp_c Sp_s [(\bar{G}^{-1}\partial_n\bar{G})
(\bar{G}^{-1}\partial_m\bar{G}) (\bar{G}^{-1}\partial_k\bar{G})]
\label{ao1}
\end{equation}
\noindent
For the tensor $\bar{B}_{nmk}$
\begin{equation}
\bar{B}_{nmk} (q) \approx \bar{B}_{nmk}(0) + O(q^{2})
\label{bnmk}
\end{equation}
\noindent
with
\begin{equation}
\bar{B}_{nmk}(0) = \int_{-\pi /a}^{\pi /a}\frac{d^{3}p}{(2\pi)^{3}} Sp
[\eta^{-}_n \bar{G}^{-1} \eta^{-}_m \bar{G}^{-1} \eta^{-}_k \bar{G}^{-1}
e^{-i(p_n+p_m+p_k)}  +  ...] = \Pi_{nmk}
\label{bnmko}
\end{equation}
\noindent
using the property
$$
\sum A_n A_m A_k \Pi_{nmk} =\frac{1}{2}\sum A_n A_m A_k(\Pi_{nmk} -\Pi_{nkm})
$$
we can find
\begin{equation}
\frac{1}{2}(\Pi_{nmk} -\Pi_{nkm}) =
-[\partial_k \bar{\Pi}_{nm}(q)] \mid_{q=0} = -a_0 \epsilon_{nmk}
\label{ao2}
\end{equation}
\noindent
Then substituting Eqs.(\ref{piexp})-(\ref{ao2}) into the definitions
of $\Gamma_2(A)$ and $\Gamma_3(A)$
we obtain (with $\overline{A} \rightarrow g \overline{A}$)
\begin{equation}
\Gamma_2(A) - \frac{2}{3} \Gamma_3(A) = S_{CS} + O(q^{2})
\label{fr}
\end{equation}
\noindent
and
\begin{equation}
k = a_0
\label{ao3}
\end{equation}
\noindent
using the definition Eq.(\ref{csmsp}).
{}From the structure of the colour traces in Eq.(\ref{ao1})
$$
Sp_c [\bar{G}^{-1}t^{a} \bar{G}^{-1}t^{b} \bar{G}^{-1}]
$$
it follows that $a_0$ is not proportional to $\Lambda$ and hence may be
nonzero when the $<A_0>$-condensate is absent. Hence, the first conclusion
from above ascertains the absence of simple perturbative
interrelation between $<A_0>$ and generation of the Chern-Simons term.
Now let us take $\Lambda = 0$ and the Green function to be proportional
to the unit matrix in colour space. Keeping the spinor  traces  in
the definition of $a_0$  and redenoting
$$
S_{CS} \rightarrow a_0 S_{CS}
$$
we finally find  for the effective action
\begin{equation}
\Gamma_{eff}(A) = \frac{a_0}{2}S_{CS} + O(q^{2}) + \tilde{\Gamma}_{eff}(A)
\label{gfa}
\end{equation}
\noindent
and for $\eta^{\pm}_0$ we have
$$
\eta^{\pm}_0 = \exp (\pm a_{\beta}\mu k) (r \pm \gamma_0).
$$
Analyzing $a_0$  we consider three possibilities.

1) $k=1; r=s \exp(i\theta \gamma_5)$. It is convenient for further
calculations to represent $\bar{G}^{-1}$ in Eq.(\ref{grf5})
in the form
\begin{equation}
\bar{G}^{-1} = \frac{\bar{G}_-}{\bar{G}_0} \ ,
\bar{G}_- = \bar{G} (-p, -\alpha, -\theta, -\mu)
\label{gnf}
\end{equation}
\noindent
where
\begin{equation}
\bar{G}_{0} = \bar{G} (p,\alpha,\theta,\mu)
\bar{G} (-p, -\alpha, -\theta, -\mu)
\label{gnot}
\end{equation}
\noindent
and contains no $\gamma$-matrices. It does not make any difficulty to
demonstrate that
\begin{equation}
a_0 =-\frac{i}{6} \int_{-\pi /a}^{\pi /a}\frac{d^{3}p}{(2\pi)^{3}}
\frac{\cos ap_n \cos ap_m \cos ap_k}{(\bar{G}_{0})^{3}} \Lambda_{nmk}
\label{ao4}
\end{equation}
\noindent
with
\begin{equation}
\Lambda_{nmk} = \epsilon_{nmk}Sp[\bar{G}_- \gamma_n \bar{G}_- \gamma_m
\bar{G}_- \gamma_k]
\label{l}
\end{equation}
\noindent
Tracing over spinor indices in Eq.(\ref{l}) we get convinced that
\begin{equation}
a_0 = 0
\label{ao5}
\end{equation}
\noindent
(for every combination of $\gamma_0\gamma_n\gamma_m\gamma_k$
there exists the combination
with an odd number of permutations resulting in a cancelation of
each other). In a sense, this result explains the one of Ref.\cite{4}. If
we were  dealing with the left-hand fermions only as in \cite{4}
it would lead in our notations to
\begin{equation}
\Lambda_{nmk} = \epsilon_{nmk}Sp[\bar{G}_- \sigma_n \bar{G}_- \sigma_m
\bar{G}_- \sigma_k]
\label{l1}
\end{equation}
and $\gamma_0 = 0$.
Then $a_0$ is expressed through the winding number of the free fermion
propagator and is still finite in the limit $m \rightarrow 0$.
We understand  in
this case that the left-hand fermion contribution annihilates
implicitly the contribution of the right-hand fermions.  Any nonzero
contribution in this case demands the symmetry to be broken.

2) $k = \gamma_5$.
In the continuum limit it corresponds to the term
$i\mu \bar{\Psi}\gamma_0 \gamma_5 \Psi$ in the Lagrangian
instead of $i\mu \bar{\Psi}\gamma_0 \Psi$ \cite{32}.
The Eqs.(\ref{gnf})-(\ref{l})
do not change and we have up to an unessential constant
\begin{equation}
a_0 \approx  i\mu \int_{-\pi /a}^{\pi /a}\frac{d^{3}p}{(2\pi)^{3}}
\frac{\cos ap_1 \cos ap_2 \cos ap_3}{(\bar{G}_{0})^{3}} \lambda^{2} =
i \mu I(m, \mu , \beta).
\label{ao6}
\end{equation}
\noindent
Futher we drop the potentials $\alpha_{\sigma}$ as they
are not important in generating $a_0$ and then
\begin{equation}
\lambda \approx  m -2\pi (\beta)^{-1} s \sin a_{\beta}\mu \sin \theta +
\frac{2s \cos \theta}{a_{\sigma}} \sum_{\sigma=1}^{d}
[\cos ap_{\sigma} - 1]
\label{lam}
\end{equation}
\noindent
Eq.(\ref{ao6}) and this definition make the difference from
the three-dimensional case and from the left-hand  fermion  model  fairly
transparent $I \sim \frac{1}{m}$ and $a_0 \rightarrow 0$ at the limit
$m \rightarrow \infty$.
Moreover, $I \sim \beta$ at $\beta \rightarrow 0$
which is in agreement with \cite{4}. It is quite evident that the
Chern-Simons term is always induced if one introduces
different chemical potentials for the left- and right-hand fermions as
in (\ref{chempot}).

3) The most interesting case is $\mu =0$ and $r=s \exp(i\theta \gamma_5)
\gamma_0$.
Its classical limit results in the initial action (\ref{stot}) with
$\mu_B =0$. However, the explicit calculation gives that
\begin{equation}
a_0 \approx \sin \theta v \frac{\partial^{2}}{\partial v^{2}} F(v) \ , \
v= (ma + 2\cos \theta)^{2}
\label{ao7}
\end{equation}
\noindent
where
\begin{equation}
F(v) = \frac{6}{(2\pi)^{3}} \int_{0}^{1} \frac{r^{4}}{v+4r^{2}}dr
\int_{0}^{\pi}d\Psi\int_{0}^{2\pi}d\phi \sin \Psi
[\sqrt{1-(r\sin \Psi \cos \phi)^{2}}-1]
\label{F}
\end{equation}
\noindent
In fact, Eq.(\ref{ao6}) coincides with the expression for the
$\theta$-term coefficient \cite{29}
(in that reference one could  also  find  its approximate
calculation). Thus, in this case the breaking of $CP$-symmetry at
the quantum level is quite enough although at the moment it is
unclear why one should take just this value of the Wilson parameter $r$.

At the end of this section we would like to discuss once more the
conclusion about universality of the abandoned part of the action.
Usually similar proofs are based on the Reisz theorem \cite{35},
like in Ref.\cite{24} where it has been successfully used  for
three-dimensional theory at zero-temperature.
Unfortunately, we were not able
to provide a finite temperature generalization of this  theorem  and
it means, strictly  speaking, that the question as to the universality
of the $\tilde{\Gamma}_{eff}(A)$ and $O(q^{2})$-corrections
is still pending.
However, if we restrict this consideration to static fields in
$O(q^{2},A)$ only, the conclusion about the independence of all
corrections of the parameter $r$ (i.e. universality)
can be drawn as it corresponds to the regime analogous to Ref.\cite{24}.

\section{Generalized Kogut-Susskind fermions}

Here we investigate lattice gauge theories on the torus with
staggered fermions described by the action
\begin{equation}
S_{K-S} = \frac{1}{2}\sum_{x,n=-d}^{d} \eta_{n}(x)\bar{\Psi}(x)
U_{n}(x)\Psi(x+n) + ma\sum_{x}\bar{\Psi}(x)\Psi(x)
\label{KSact}
\end{equation}
\noindent
and we use the following rules
\begin {equation}
\eta_{-n}(x) = -\eta_n(x); \ a_{-n}=-a_n; \ U_{-n}(x)=U^{+}_n(x-n)
\label{prop}
\end {equation}
\noindent
It is known from the original paper by Susskind \cite{36} that
the $\eta$ symbol in Eq.(\ref{KSact}) arises in the action after
diagonalization of the initial naive action over spinor indices. As
a diagonalizing operator one takes usually the  unitary operator
\begin{equation}
T_0 = \prod_{\nu=1}^{d} (\gamma_{\nu})^{x_{\nu}}
\label{dop1}
\end{equation}
\noindent
and then the $\eta$ symbol is in the well-known standard form
\begin{equation}
\eta_n^{0}(x) = (-1)^{x_{1}+x_{2}+...+x_{n-1}}
\label{eta1}
\end{equation}
\noindent
It is quite understandable that (\ref{dop1}) is not the unique choice for
this operator, even in (\ref{eta1}) it is defined up to a factor
$\epsilon$ for which $\epsilon^{2} = 1$. Taking , for example,
$\epsilon=\exp [i \pi k(x)]$ where $k(x)$ is any integer we
obtain a more general form of the unitary diagonalization operator
\begin{equation}
T_1 = [\prod_{\nu=1}^{d} (\gamma_{\nu})^{x_{\nu}}] e^{i\pi k(x)}
\label{dop2}
\end{equation}
\noindent
and for the $\eta$ symbol we have
\begin{equation}
\eta_n(x) = T_1^{+}(x) \gamma_n T_1(x+n) =
\eta_n^{0}(x) \exp(i\pi[k(x+n)-k(x)])
\label{eta2}
\end{equation}
\noindent
However, at any choice of $\eta$ the action (\ref{KSact}) in the continuum
limit describes massive Dirac fermions with four flavours.
As to the chiral properties of Eq.(\ref{KSact}), in massless limit it is
invariant under global $U(N_f)\otimes U(N_f)$
transformations of fermionic fields \cite{37}.
At the first sight the theories with different form of $\eta$ should be
equivalent but this is not always the case. To our mind there exists,
at least, one choice for $\eta$ when the fermionic determinant can
generate a nonuniversal contribution to an effective action. In a
sense it looks like the situation with the above discussed Wilson
parameter $r$ subdividing the original classical theory into
different universality classes related to a concrete choice of
$x$-dependence of integer $k$ in Eq.(\ref{eta2}). We are able to
demonstrate  that  nonuniversal contribution for the gauge theory on
torus at rather plausible assumptions should contain the Chern-Simons
action, although we cannot prove that it is the unique one.

We start from the four-dimensional fermionic determinant and have
for Eq.(\ref{geff})
\begin{equation}
\Gamma_{eff} = -N \ln ma - \sum_{l}\frac{(-1)^{l}}{l(2ma)^{l}}
Sp [(D)^{l}]
\label{geffks1}
\end{equation}
\noindent
with
\begin{equation}
D = \sum_{n=-d}^{d}D^{n} \ , \ D^{n}_{xy}=\delta_{x}^{y+n}\eta_{n}(x)
U_n(x)e^{-\mu a}
\label{notks}
\end{equation}
\noindent
The closed loops only will contribute to $\Gamma_{eff}$ just due to the
original gauge invariance and the $l$-th order in the expansion
(\ref{geffks1}) corresponds to the loop of length $l$ as
\begin{equation}
I_l = Sp [(D)^{l}] = \sum_{x_j,n_j;x_{l+1}=x_1}
Sp[\prod_{j=1}^{l} \delta_{x_j}^{x_j+n_j} \eta_{n_j}(x_j)
U_{n_j}(x_j)] \ f(\mu)
\label{il1}
\end{equation}
\noindent
with $f(\mu)=1$ if the loop does not wind the whole torus and
$f(\mu) = \exp(-\beta \mu) \ (\exp(\beta \mu))$
when the loop winds the torus in the positive (negative) direction.
Summing up in Eq.(\ref{il1}) we find
\begin{eqnarray}
I_l &=& \sum_{x_j,n_j;x_{l+1}=x_1} Sp_c f(\mu)
[\eta_{n_1} \eta_{n_2} ... \eta_{n_l} U_{n_1}(x)U_{n_2}(x+n_1) ...U_{n_l}(x)]
		\nonumber   \\
 &=& \sum_{C}f_{C}(\mu)\eta (C)SpU(C)
\label{il2}
\end{eqnarray}
\noindent
where $C$ is a closed loop.
Here we used the following rules
\begin{equation}
\eta_n \eta_m = \eta_n(x) \eta_m(x+n)
\label{eta3}
\end{equation}
\noindent
and keeping it in mind we are able to show the existence of
$\eta$ symbols obeying (in particular, like Eq.(\ref{eta1}))
\begin{equation}
\{ \eta_n \eta_m \} = 2 \delta_{nm}
\label{et}
\end{equation}
\noindent
Our strategy is to demonstrate, firstly, that Eq.(\ref{il2}) includes a
contribution transforming to $\theta$-term in the continuum limit.
We need then $\eta$-symbols with the following properties
\begin{equation}
\eta_n \eta_m \eta_k \eta_l = \epsilon_{nmkl}
\label{et4}
\end{equation}
\noindent
for $n \neq m \neq k \neq l$
and for any arbitrary indices the contributions like $\eta_n \eta_m$ appear.
Putting down $k(x) = x_1 x_3$ in (\ref{dop2}) we find
\begin{eqnarray}
T = T_0 (-1)^{x_1 x_3},    \nonumber  \\
\eta_1= (-1)^{x_3} \ , \ \eta_2= (-1)^{x_1} \ , \ \eta_3= (-1)^{x_1} \ , \
\eta_4= (-1)^{x_1+x_2+x_3}
\label{dopeta}
\end{eqnarray}
\noindent
which results in
\begin{equation}
\eta_n \eta_m \eta_k \eta_l = \epsilon_{nmkl} +\delta_{nm}\eta_k \eta_l ...
\label{etpr}
\end{equation}
\noindent
The similar $\eta$-symbols had been originally introduced in Ref.\cite{38}.

Now we consider an arbitrary $C$ in Eq.(\ref{il2}) winding around a
torus and contributing
\begin{equation}
\tilde{B}(C) = \eta_n \eta_m \eta_k \eta_l U_n(y)U_m(y+n)U_k(y+n+m)
U_l(y+n+m+k)
\label{bc}
\end{equation}
\noindent
with $n \neq m \neq k \neq l$.
Parametrizing $U_n(y)$ in the standard form with potential behaving
smoothly
\begin{equation}
U_n(y) = P\exp [ ig \int_{y}^{y+n}d\xi A_n(\xi) ] \ \in SU(N)
\label{upl}
\end{equation}
\noindent
where  $A_n = A_n^{a}t^{a}$
we get in continuum limit, expanding $U_n$ around $y$ from the
part of loop $\tilde{C}$ including four links
\begin{equation}
\tilde{B}(\tilde{C}) \rightarrow_{a \rightarrow 0} a^{4}g^{2} \epsilon_{nmkl}
\partial_n [A_m \partial_k A_l +\frac{2g}{3} A_m A_k A_l].
\label{bc1}
\end{equation}
\noindent
For $B(C) \eta (C)$ it gives
\begin{equation}
B(C) \eta (C) \rightarrow_{a \rightarrow 0} \tilde{B}(\tilde{C})
\eta (C/ \tilde{C}) [1+ iag A_p(y+n+m+k+l) ... ].
\label{bc2}
\end{equation}
\noindent
Since we need the limit of the lowest orders in $a$  and $g$
only, it is enough to consider the loops containing one
antisymmetric tensor or two ones if four links which should be covered
in the opposite direction (regarding to the initial one)
are running successively. The point is that for getting the proper
continuum limit one should take the "empty" links (i.e. number one
in the expansions such as (\ref{bc2})) as the factors and such contributions
cancel out due to the evident property
$$
\sum_{nm} \epsilon_{nmkl} =0
$$
at those "empty" links. Let us consider the case of two
antisymmetric factors available
$$
[a^{4} \epsilon_{n_1 m_1 k_1 l_1}
\partial_{n_1}(A_{m_1} \partial_{k_1} A_{l_1} +
\frac{2g}{3} A_{m_1} A_{k_1} A_{l_1})] \ 1 \ 1 ...
$$
\noindent
\begin{equation}
[ a^{4} \epsilon_{n_2 m_2 k_2 l_2}
\partial_{n_2} (A_{m_2} \partial_{k_2} A_{l_2} +
\frac{2g}{3} A_{m_2} A_{k_2} A_{l_2})] \ 1 \ 1 ...
\label{2eps}
\end{equation}
\noindent
The boundary conditions for torus of period $l$  take the
following form \cite{15}
\begin{equation}
A_{\mu}(x)= \Omega_{\nu}^{-1}(x) A_{\mu}(x+l_{\nu}) \Omega_{\nu}(x) -
[\partial_{\mu} \Omega_{\nu}^{-1}(x)] \Omega_{\nu}(x)
\label{tbc}
\end{equation}
\noindent
where $\Omega_{\nu}(x) \in SU(N)$ and satisfies the conditions \cite{16}
\begin{equation}
\Omega_{\mu}(x+l_{\nu}) \Omega_{\nu}(x) = \Omega_{\nu}(x+l_{\mu})
\Omega_{\mu}(x)
\label{tbc1}
\end{equation}
\noindent
when the matter fields are present. The matrices $\Omega(x)$
subdivide the theory into different topologically nonequivalent sectors
which one should sum over (this is equivalent to summing over all boundary
conditions \cite{16}). Substituting
\begin{equation}
A_{\mu}(x) = \Omega^{-1}(x) A_{\mu}(x+l) \Omega(x) - [\partial_{\mu}
\Omega^{-1}(x)] \Omega(x)
\label{tgtr}
\end{equation}
\noindent
into Eq.(\ref{2eps}) and summing over similar loops in Eq.(\ref{il2}) we have
in the continuum limit
\begin{equation}
I_l \approx \alpha n \int d^{4}x F_{\mu \nu} \tilde{F}_{\mu \nu}= \alpha n \int
d^{4}x \epsilon_{\mu \nu \rho \sigma} Sp[\partial_{\mu}(A_{\nu} \partial_{\rho}
A_{\sigma} + \frac{2g}{3} A_{\nu} A_{\rho} A_{\sigma})]
\label{il3}
\end{equation}
\noindent
with
\begin{equation}
n = \frac{1}{24 \pi^{2}} \int ds_{\mu} \epsilon_{\mu \nu \alpha \beta}
Sp(V_{\nu}V_{\alpha}V_{\beta})
\label{wnt}
\end{equation}
\noindent
where $V_{\nu} = \Omega \partial_{\nu} \Omega^{-1}$.

Considering the loops with one antisymmetric tensor, one should
keep in mind that the loops which differ in a  couple  of "empty"
links only, for instance, $\eta_n(x) \eta_m(x+n)$ and
$\eta_m(x) \eta_n(x+m)$ cancel
out in the continunm limit, owing to Eq.(\ref{et}), when summing over all
loops. It follows also from (\ref{et}) that only those loops survive
where one makes two steps successively in each direction for any
"empty" link or, in other words, only loops with even number of steps
over the "empty" links to each direction. More comprehensive analysis
shows that an odd number of steps is admissible in any (or in any
but one) direction. Thus, the total link number in each or in each but
one direction must be an odd number if one
would like the $\theta$-term to appear. Then, however, neither the
operator $T_0$ in Eq.(\ref{dop1}) nor the operator $T$ in Eq.(\ref{dop2})
diagonalizes the initial fermionic matrix $D = \gamma_nD^{n}$
placed between the first and the
$(N-1)$-th sites at the last link on torus. We do not know any
operator of (\ref{dopeta}) or any others which could be suitable for a lattice
of odd length. That is why we consider in what follows the loops
which differ in a couple of "empty" links only. In  this
situation the $\theta$-term is still available only when a
nonperturbative vacuum does exist in the theory. We mean the
following.

Let us take the matrix $U_n(x)$ in Eq.(\ref{upl}) as
\begin{equation}
U_n(x) = Z_n(x)\tilde{U}_n(x); \  Z \in Z(N) ,
\tilde{U} \in \frac{SU(N)}{Z(N)}
\label{upld}
\end{equation}
Then the contribution of the loop parts which differ in a
couple of links is
\begin{equation}
\eta_n \eta_m U_n(x) U_m(x) + \eta_m \eta_n U_m(x) U_n(x+m) \approx
1 - Z(p) \tilde{U}(p)
\label{ucl}
\end{equation}
\noindent
($p$ is the plaquette).
The following reasonings are applicable not only for the theory on
torus but for any theory with $\Omega = R^{3} \otimes S^{1}$.
Two options are possible at least.

(i) $Z(N)$-mechanism of quark confinement where the vortex condensate
changes the vacuum structure and the symmetry $SU(N) \rightarrow Z(N)$
is well elaborated \cite{39}. Then, unlike the
naive expectation, the important contribution to the path
integral comes from not only an identity element of group but from
all elements of $Z(N)$. In this example it means that we must not
put $Z(p)\tilde{U}(p) = 1$ into Eq.(\ref{ucl}) if we need to
reproduce properly the structure of quantum theory vacuum.
It seems to us that the more reasonable way is to sum over all
configurations of center of group in the path integral and
then to construct the continuum limit of $\tilde{U}(p)$.
Unfortunately, we have not managed to perform these operations,
although several promising hints could be found in \cite{17} to succeed.
Nevertheless, it
is clear that the $\theta$-term may be generated in this way and if the
vortex condensate appears in such a theory \cite{40} the coefficient of
$\theta$-term should be proportional to that condensate. Thus, the
fermionic determinant could generate the $\theta$-term with a
coefficient depending on a winding number as in
Eqs.(\ref{il3}),(\ref{wnt}) or on the configurations related to $Z(N)$-group.
In the deconfined phase at high temperature where the field
configurations
\begin{equation}
A_0 = const, \ \partial_{0}A_n(x) \rightarrow 0
\label{cond}
\end{equation}
\noindent
are dominating, the $\theta$-term is reduced to the Chern-Simons action
with the coefficient $2n \beta <A_0>$ \cite{41}.

(ii) Treating again the theory given by Eq.(\ref{KSact}) with $\eta$ symbols
(\ref{dopeta}) we fix the static diagonal gauge $\partial_0 A_0 = 0$ and
following \cite{42} we reduce the fermionic determinant to the effective three
dimentional theory. Specific feature of that operation is an
averaging of lattice gauge field matrices but not of the field
$A_{\mu}(x)$. It is provided by the following change of
variables in the path integral
\begin{equation}
U_n(\vec{x},t) \rightarrow \Phi_n(\vec{x}) =
\frac{1}{N_{\beta}} \sum_{t=1}^{N_{\beta}} U_n(\vec{x},t)
\label{phi}
\end{equation}
\noindent
where $\Phi_n(\vec{x})$ can be given as
\begin{equation}
\Phi_n(\vec{x}) = \rho_n(\vec{x}) \overline{U}_n(\vec{x})
\label{rho}
\end{equation}
\noindent
with $\overline{U}_n(\vec{x}) \in SU(N)$ and $\rho_n$ being a positively
defined matrix for $SU(N), N \geq 3$ and $0 \leq \rho \leq 1$ for $SU(2)$.
The theory obtained is invariant with respect to time-independent gauge
transformations. Eq.(\ref{phi})
turns into standard reduction procedure with $\rho_n=1$ in the naive
continuum limit and means that taking high-order expansion
leads to appearance of a new vector field.
Moreover, this consideration could be done explicitly if one could
calculate the corresponding Jacobian. Omitting the pure gauge
part which is not interesting now, for $SU(2)$ theory
the Jacobian takes the form \cite{42}
\begin{equation}
J \sim \exp [-N_{\beta} \rho_n^{2}(x)] \ \mbox{at} \ N_{\beta} \gg 1
\label{jac}
\end{equation}
\noindent
Conceptually it is rather important that the field $\rho_n$ may be
tractable as a dielectric field \cite{18}. Moreover, the matrices $U_n$
are parametrized by smooth potentials $A_n(x)$ as in Eq.(\ref{upl}) and
$<\rho_n> = 0$ for the pure gauge theory  and  $<\rho_n> = const$ for
the theory with fermions (in the confined phase the field
$\rho_n(x)$ concentrates into a squeezed flux tube
of the group centre charges) \cite{18}, \cite{42}.

Implying massless quarks we have the fermion determinant
reduced as
\begin{equation}
Det[1+\frac{\beta}{a_{\sigma}}R_{\vec{x}}\sum_{n}\bar{D}^{n} + h.c.]
\prod_{x} Det_{c} R^{-1}_{\vec{x}}
\label{detks}
\end{equation}
\noindent
and the effective action, expanding in $\beta$, as
\begin{equation}
\Gamma_{eff} = -\sum_{\vec{x}} \ln Det_{c} R^{-1}_{\vec{x}} +
\frac{1}{2}\sum_{l}(\frac{\beta}{a_{\sigma}})^{2l} \frac{1}{l}
Sp[(\bar{D})^{2l}]
\label{5}
\end{equation}
\noindent
where we have already used that only closed (i.e. with even number
of links) loops contribute and
\begin{equation}
\bar{D} = R_{\vec{x}}\sum_{n=-d}^{d}\bar{D}^{n} + h.c.
\label{6}
\end{equation}
\noindent
The following notations have been introduced
\begin{equation}
R^{ij}_{\vec{x}} = (1 + W^{ij}_{\vec{x}})^{-1}
\label{r}
\end{equation}
\noindent
\begin{equation}
\bar{D}^{n}_{\vec{x} \vec{y}} =
\bar{\eta}_n(\vec{x}) \rho_n(\vec{x}) \bar{U}_n(\vec{x})
\delta_{\vec{y}}^{\vec{x}+n}
\label{Dr}
\end{equation}
and $W_{\vec{x}}^{ij}$ are the components of the Polyakov loop
\begin{equation}
W_{\vec{x}} = \exp (ig \beta A_0 (\vec{x}))
\label{Wl}
\end{equation}
\noindent
A chemical potential is present according to the change
$W \rightarrow \exp (-\beta \mu) W$, \
$W^{\ast} \rightarrow \exp (\beta \mu) W^{\ast} $
and the "reduced"  $\bar{\eta}$ symbols are
\begin{equation}
\bar{\eta}_1 = (-1)^{x_1 + x_2}, \ \bar{\eta}_2 = (-1)^{x_2 + x_3},
\ \bar{\eta}_3 = (-1)^{x_3 + x_1}
\label{reta}
\end{equation}
\noindent
The property of Eq.(\ref{et}) is automatically satisfied and
\begin{equation}
\bar{\eta}_n \bar{\eta}_m \bar{\eta}_k = \epsilon_{nmk}
\label{reta1}
\end{equation}
\noindent
for $n \neq m \neq k$ taking into account (\ref{eta3}).
Going to the continuum limit
we imply that we have to use the complete form of the Polyakov loop (\ref{Wl})
in the limit i.e. we do not expand it around unit matrix
unlike other gauge configurations $\overline{U}_n(x)$.
Admitting the configurations
\begin{equation}
SpW_{\vec{x}} \approx <SpW>, \  \rho_n(\vec{x}) \approx \ <\rho>
\label{bascon}
\end{equation}
\noindent
to be the basic ones we regard in (\ref{5}) the loops with
one antisymmetric tensor coming from three succesive links
$\eta_n \eta_m \eta_k$. In the continuum limit the smooth potentials of those
links generate just the Chern-Simons action
\begin{equation}
a^{3} \epsilon_{nmk} (A_n\partial_m A_k + \frac{2}{3}A_nA_mA_k)
\rho_n \rho_m \rho_k.
\label{cscl}
\end{equation}
\noindent
The remaining links of such loops provide the factors $W\rho_nW$.  Summing
over loops distinguishing a couple of these factors we find their sum
to be proportional to the strength tensor of the dielectric field
$F_{nm}(\rho)$.
This contribution should be added by the contribution of such loops but going
the opposite direction. Then Eq.(\ref{cscl}) acquires the opposite sign and
the change $W \rightarrow W^{\ast}$ should be done. Assuming
$<SpW> \in Z(N)$ let us expand $R_{\vec{x}}$ in Eq.(\ref{r}) as
\begin{equation}
R_{\vec{x}}=W_0 + i \lambda^{3}W^{3} + i \lambda^{8}W^{8}
\label{r3}
\end{equation}
\noindent
where
\begin{equation}
3W_0 = Sp R, \ 3W^{a} = Sp (\lambda^{a}R).
\label{r8}
\end{equation}
\noindent
Then, as $W \in Z(N)$ all $W$ commute with the field $A$ and it is easy
to understand that the sum of two loops going the opposite
ways is proportional to the imaginary part of the Polyakov
line. If the condensate $<A_0>$ falls the calculations
are much more complicated but the Chern-Simons action, nevertheless, is
proportional to $ImW$ with redefinition of gauge potentials as in
\cite{41}. Thus, the effective action (\ref{5}) provides the
contribution
\begin{equation}
i \alpha_0 <ImW> F_{mn}(\rho) \int d^{3}x L_{CS}
\label{100}
\end{equation}
\noindent
The coefficient $\alpha_0$ absorbs, firstly, the factors $a_l<\rho>^{l}$
($l$ is the loop of length $l$ and $a_l$ is its "weight");
secondly, the contribution of loops with several antisymmetric
tensors as at $<\rho> \neq 0$ they do not cancel; thirdly, the octet part
dependence of the Polyakov loop $W^{a}$. As to an exact calculation
of $\alpha_0$ it can be estimated for the loops of small sizes only.
This result, however, is not valid in $SU(2)$ gauge theory because
$SpIm W = 0$
and $F_{mn}(<\rho>) = 0$ as $\rho_n$ is an abelian field in that case. The
field $\rho_n(x)$ may be nonabelian for $SU(3)$ theory \cite{43} and if
$\rho_n(x)$ belongs to the diagonal subgroup one could demand
stronger condition $F_{mn}(\rho) \neq 0$ to be fulfilled
(like for $SU(2)$ as well).
Concerning the conditions of Eq.(\ref{bascon}) they are quite relevant for the
deconfined phase where $<SpW> \neq 0$ and  the dominant contribution
comes from any configuration $W = \exp (2 \pi i n / 3 ),\ n=0, \ \pm 1$
and its fluctuations due to global symmetry breaking. If $n = \pm 1$
then $<Im W> \neq 0$  leading to $C$ (or $CP$)-symmetry
breaking \cite{19} and the Chern-Simons term generation. If $n = 0$, the
loops going the opposite directions cancel each other. But even in
this case the baryonic chemical potential attaches
proper different weight factors to those loops and their sum is
proportional to $\sinh \mu_B \beta$. It suggests that the
following substitution
$$
i<Im W> \rightarrow i<Im W> \cosh \mu_B \beta + <Re W> \sinh \mu_B \beta
$$
should be done in Eq.(\ref{100}).

\section{Summary}

Actually in the Introduction of the present paper we put,
in essense, two principal questions:

i) could the high temperature QCD belong to the universality class
with nontrivial parameter $r$ ? Eventually it seems to us that
small (dynamical or spontaneous) violation of the $CP$-symmetry
should be present in the theory. The Chern-Simons action appears
then automatically which we have demonstrated above;

ii) is it still possible to generate the Chern-Simons term in
effective theory when the parameter $r$ is trivial (for example, $\theta =
0$ and $0 < s \leq 1$). It follows from our results that $S_{CS}$ can arise
here also but except for the $CP$-symmetry breakdown when we need
to have additional constraints for chemical potentials of the theory.

Now we could answer these questions qualitatively concluding with:

(i) the Chern-Simons action can be perturbatively generated
by the fermionic determinant (dealing with generalized Wilson
fermions on a lattice) when the Wilson parameter $r = 1$ and the
baryonic chemical potential is introduced in the form
$i\mu \bar{\Psi}\gamma_0 \gamma_5 \Psi$,
which corresponds, as a matter of fact, to different chemical
potentials for the left-hand and the right-hand fermions. In a sense we
disrecovered this result in more proper calculations as it was
obtained earlier in Ref.\cite{32} with fixing $A_0 = 0$. Besides, the
parameter $r$ may also take the form
$r = s \exp (i \theta \gamma_5 ) \gamma_0$ with an
arbitrary chemical potentials (including $\mu_B= 0$). $CP$-symmetry is
obviously broken with this $r$ but a classical limit is going to the
standard continuum theory and a nonuniversal contribution could take the
form of the Chern-Simons action.

(ii) dealing with the Kogut-Susskind fermions, the $\theta$-term may be
generated either by boundary conditions of torus or by singular
potentials related to the centre of the gauge group. At high
temperatures when the $<A_0>$-condensate is available the $\theta$-term is
reduced to the Chern-Simons action.

(iii) in the $SU(3)$ gauge theory with nonperturbative
dielectric vacuum appearing in the reduced theory, the Chern-Simons
action is generated when the expectation value of the imaginary part
of the Polyakov line is non-zero or baryonic chemical potential does
exist. The latter takes place when $<A_0^{3}>$ and $<A_0^{8}>$
are nonzero in the deconfined phase
or $<Sp W> = \exp (\pm i2 \pi /3 )$.

For both items (ii) and (iii) we have to use the general form of the
$\eta$ symbols for the Kogut-Susskind fermions with special properties
to get the desirable result.

The authors are grateful many colleagues for fruitful discussions
but the remarks of Yu. Makeenko and G.Semenoff who had drawn our
attention to the Rutherford's paper \cite{32} were especially valuable.

\end{document}